\documentclass[%
 reprint,
 amsmath,amssymb,
 aps,
]{revtex4-2}

\usepackage{comment}
\usepackage[normalem]{ulem}
\usepackage{graphicx}
\usepackage{dcolumn}
\usepackage{bm}
\usepackage{hyperref}

\usepackage{soul}
\usepackage[dvipsnames]{xcolor}
\setstcolor{red}

\begin{document}

\title{Universal and non-universal signatures in the scaling functions of critical variables}

\author{Gianluca Teza}%
\affiliation{Max Planck Institute for the Physics of Complex Systems, Nöthnitzer Strasse 38, 01187 Dresden, Germany}
\affiliation{Department of Physics of Complex Systems, Weizmann Institute of Science, Rehovot 7610001, Israel}
\author{Attilio L. Stella}
\email{stella@pd.infn.it}
\affiliation{Department of Physics and Astronomy, University of Padova, Via Marzolo 8, I-35131 Padova, Italy}
\affiliation{INFN, Sezione di Padova, Via Marzolo 8, I-35131 Padova, Italy}
\date{\today}

\begin{abstract}
The view that the probability density function (PDF) of a key statistical variable, anomalously scaled by size or time, could furnish a hallmark of universal behavior contrasts with the circumstance that such density sensibly depends on non-universal features. We solve this apparent contradiction by demonstrating that both non-universal amplitudes and universal exponents of leading critical singularities in large deviation functions are determined by the PDF tails, whose form is argued on extensivity. This unexplored scenario implies a universal form of central limit theorem at criticality and is confirmed by exact calculations for mean field Ising models in equilibrium and for anomalous diffusion models.
\end{abstract}

\maketitle

The confirmation of scaling and universality, together with the calculation of critical exponents on which one can test such universality, are among the most fundamental achievements of the theory of equilibrium critical phenomena \cite{goldenfeld2018lectures,cardy1996scaling,kadanoff2000statistical}.
Much progress in this field was possible through finite size scaling techniques, which allowed to determine critical exponents by extrapolating to the thermodynamic limit finite systems properties \cite{privman1990finite,cardy2012finite,Privman1984}.
An interesting prerogative of finite systems at criticality is also the fact that the probability distribution of the order parameter, if properly scaled, converges to a well defined scaling PDF \cite{binder1981,nicolaides1988universal}.
For example, in an Ising model with $N$ spins in $d>1$ dimensions, the probability distribution of the magnetization $M$ at the critical temperature $T_c$ and in zero magnetic field is expected to converge for increasing $N$ to a limit
\begin{equation}\label{eq.scaling0}
N^{y_H/d}P_N(M,T_c) \xrightarrow[N \to \infty]~ f(M/N^{y_H/d})    
\end{equation}
with $f$ a scaling function of the rescaled magnetization $m=M/N^{y_H/d}$, and $y_H$ the magnetic Kadanoff exponent \cite{kadanoff2000statistical}.

Scaling functions were proposed as possible locations of universal signatures, with a conjectured stretched exponential decay $f(m)\sim \exp(-c |m|^{\delta+1})$ for large $|m|$ \cite{Bruce1995,Hilfer1995} modulated by the equation of state exponent $\delta$  \cite{kadanoff2000statistical} and some unknown coefficient $c>0$.
In spite of many attempts, numerical analyses of $f$ proved to be extremely difficult and far from conclusive for the tails \cite{tsypin2000probability,hilfer2003multicanonical,hilfer2005multicanonical,malakis2006universal} (for rigorous proofs in 2D see \cite{mccoy1973two,camia2016planar}).
Analytical evidence of this type of decay, with precisely the form conjectured for the Ising magnetization scaling function \cite{Bruce1995,Hilfer1995}  \footnote{This form also included a power law factor multiplying the stretched exponential. However, this factor does not influence extensive quantities.}, was most recently found in a series of paradigmatic models of anomalous diffusion \cite{stella2023anomalous,stella2023universal} with time and displacement playing the roles of system's size and magnetization, respectively.
Such decay was also identified as responsible of power law singularities in the large deviation functions of displacement, suggesting the importance of these models for proving properties out of reach in the Ising case.

The probabilistic interpretation of the renormalization group \cite{Jona1975,jona2001renormalization} suggested to regard non-Gaussian scaling functions as solutions of stability conditions for the PDF's of sums of strongly correlated random variables \cite{cassandro1978critical,gallavotti1975}.
It remains an open issue to identify up to what extent these shapes are reflecting universal (or non-universal \footnote{Typical non-universal quantities are the amplitudes of the power law singularities, which can be of key importance also in the numerical or experimental identification of the exponents themselves.}) properties, as $f$ strongly depends on features which can be different within the same universality class, such as the boundary conditions of the finite systems from which it is extrapolated \cite{binder1981,antal2004,kaneda2001finite}.
The absence of precise criteria for identifying universal and non-universal features of scaling functions prevented so far the formulation of general forms of limit theorems at equilibrium criticality, as well as a thorough investigation of critical power-law singularities in the rate functions of large deviation theory  \cite{Touchette2009,touchette2013,balog2022critical}. 

In this work, we focus on the magnetization of the critical Ising model in equilibrium as a paradigmatic example of variable obeying anomalous scaling.
We demonstrate that the extensivity of an auxiliary form of cumulant generator determines the stretched exponential decay of the scaling function conjectured for the Ising model.
This decay is shown to account for both universal exponents and non-universal amplitudes of leading singularities in all large deviation functions, opening to a universal generalization of the central limit theorem at criticality.
The scenario is exactly confirmed in the case of mean field interactions and for the displacement in time of anomalous diffusion models.


We start by considering the example of an Ising model in equilibrium.
In $d=2$ or $d=3$, the system undergoes a phase transition at a finite critical temperature $T=T_c>0$ \cite{kadanoff2000statistical}.
We consider a box with $N$ spins, e.g. on a square or cubic lattice, and with open boundary conditions.
The same argument applies to different lattices or periodic boundary conditions.
Given a generic configuration $\vec{\sigma}=\{\sigma_1,\sigma_2,\dots,\sigma_N\}$ of the system, where $\sigma_i=\pm 1$ is the spin at site $i=1,2,\dots,N$, the total magnetization is $M=\sum_i \sigma_i$.
The system has reduced Hamiltonian $\mathcal{H}(\vec{\sigma})=\frac{J}{k_BT} \sum_{\left<i,j\right>} \sigma_i \sigma_j +hM$.
Here $J>0$ is a ferromagnetic coupling and the sum is extended to all pairs of nearest neighbor sites, $k_B$ is the Boltzmann's constant and $h\geq 0$ is a dimensionless magnetic field (magnetic spin energy divided by $k_BT$; the same analysis can be repeated for negative $h$).
The reduced Helmoltz free energy is given by $\mathcal{F}_N(h,T) =\log(\sum_{\vec{\sigma}} e^{\mathcal{H}(\vec{\sigma})})$ where the sum is over all $2^N$ spin configurations and the following identity holds (see the Supplemental Material \cite{SM}):
\begin{equation}\label{eq.identity}
G_N(h,T):=\sum_{M}e^{hM}P_N(M,T) = e^{\mathcal{F}_N(h,T)-\mathcal{F}_N(0,T)}.
\end{equation}
Here, $P_N(M,T)$ is the probability of observing the system in a generic configuration with total magnetization $M$ for $h=0$.
Thus, $P_N(M,T_c)$ refers to the system becoming critical in the $N\to\infty$ thermodynamic limit.
The function $\log G_N $ is the generator of the cumulants of this probability distribution $P$ by derivation with respect to $h$ at $h=0$.

If we set $T=T_c$ in Eq. \ref{eq.identity} for large $N$ the right hand side should grow as $\sim e^{ N[g(h,T_c)-g(0,T_c)]+\dots}$, where $g$ is the free energy per site in the thermodynamic limit (the dots represent sub-extensive terms that depend on the specific choice of boundary conditions \footnote{Surface terms or edge free energies for open boundary conditions lead to contributions proportional to $N^{(d-1)/d}$ or $N^{(d-2)/d}$, while corners yield terms $\propto \log N $.
These corrections are absent for periodic boundary conditions, while an $N$-independent Privman-Fisher coefficient \cite{Privman1984}, possibly dependent on boundary conditions, should be always present.}).
In large deviations language  $g(h,T_c)-g(0,T_c)$ is the scaled cumulant generating function (SCGF) \cite{Touchette2009} of the observable $M$ at $T=T_c$, with $h$ playing the role of a Laplace parameter.
We expect $g$ to be singular at criticality for $h\to 0$ \cite{kadanoff2000statistical}, so that we can write
\begin{equation}\label{eq.identity'}
    g(h,T_c)-g(0,T_c)=\lim_{N\to\infty} \frac{\log G_N(h,T_c)}{N}=A h^{d/y_H} +\dots
\end{equation}
where the first term on the r.h.s. is the leading singular term with amplitude $A$ and the dots indicate other less singular or regular terms in a whole expansion (sub-extensive terms in Eq. \ref{eq.identity} are omitted).
The determination of the leading singular term, with the magnetic Kadanoff exponent $y_H<d$, is a task which can be faced only by exact solutions (in 2D \cite{mccoy1973two}) and in most cases is more or less satisfactorily achieved by approximate numerical or renormalization group methods \cite{privman1990finite,goldenfeld2018lectures,kadanoff2000statistical}.

At criticality, $P_N(M,T_c)$ is expected to satisfy scaling as $N\to\infty$, meaning that in the thermodynamic limit this distribution translates into a PDF $f(m)$ (up to a factor $1/2$, see \cite{SM}) of the continuous argument $m=M/N^{y_H/d} \in [-\infty,+\infty]$.
Here we want to show that the elusive behavior of $f(m)$ for large $|m|$ can be obtained by studying the role of this function in characterizing the limit in Eq. \ref{eq.identity'}.
At the same time, we will demonstrate that this insight is sufficient to fully characterize the correct exponent and amplitude of the leading singularity of $g$.
Quite remarkably, given the value of $y_H/d$, the form of the asymptotic behavior of $f$ can be argued simply on the basis of the extensivity of the cumulant generator (the power of $N$ normalizing $\log G_N$ in Eq. \ref{eq.identity'}).
This extensivity in equilibrium is determined by the existence of the thermodynamic limit for the free energy density.

To accomplish all this, one has to replace $G$ by an auxiliary expression defined as
\begin{equation} \label{eq.integral}
    \bar{G}(h N^{y_H/d},T_c) =\frac{1}{2}\int_{-\infty}^{+\infty} dm~ e^{m h N^{y_H/d}}f(m)
\end{equation}
which, by differentiation with respect to $h$ at $h=0$, generates the leading large $N$ behavior of the moments of $P_N(M,T_c)$.
As we show below, $\bar{G}$ cannot be expected to reproduce the full large $N$ behavior (inclusive of sub-leading extensive terms) on the r.h.s. of Eq. \ref{eq.identity'}.
Indeed, $\bar{G}$ is a function of the single argument $h N^{y_H/d}$, instead of $h$ and $N$ separately as $G_N$.
Still, $\bar{G}$ offers the advantage to fully account for the leading singularity just by inspection of its $N \to \infty$ behavior.
At the same time, this generator allows to link this singularity to the tail of $f$. 

Assuming rapid decay of $f$ at large $|m|$, the integral in Eq. \ref{eq.integral} for large $h N^{y_H/d}$ is dominated by the maximum value of the integrand.
For $h>0$, the point of maximum $\bar{m}$ of the integrand approaches $+\infty$ in the limit $N\to\infty$.
Writing $f(m) =e^{-r(m)}$, $\bar{m}$ formally satisfies the differential equation
\begin{equation} \label{eq.mbar}
    r'(\bar{m})=hN^{y_H/d}
\end{equation}
such that the leading contribution to the logarithm of the auxiliary function is
\begin{equation} \label{eq.leading}
    \log \bar{G}(h N^{y_H/d},T_c) = hN^{y_H/d} \bar{m} - r(\bar{m})
\end{equation}
Now, if this leading contribution is assumed to be $\propto N$ and $y_H/d<1$, from the last two equations follows that $\bar{m}\propto N^{1-y_H/d}$ and $r(\bar{m})\propto N$.
This necessarily implies that $r$ must be proportional to a power of $\bar{m}$.
Indeed, $r(\bar{m})$ is expected to grow as $\bar{m} \to +\infty$, but a growth faster than a power, e.g. exponential, would lead to $\bar{m} \sim \log(hN^{y_H/d})$, not fulfilling the required extensivity.
Following the notations adopted in Refs. \cite{Bruce1995,hilfer1994}, we write $r(\bar{m})= c \bar{m}^{\delta+1}$ and, consistently,
\begin{equation} \label{eq.mbar1}
    \bar{m}= \left( \frac{h N^{y_H/d}}{c(1+\delta)}\right)^{1/\delta}.
\end{equation}
These positions are clearly compatible with Eq. \ref{eq.mbar} and, guarantee a leading Laplace contribution $\propto N$ to the logarithm of the integral in Eq. \ref{eq.integral} with
\begin{equation}\label{eq:delta_yH}
    \delta= \frac{y_H/d}{1-y_H/d}
\end{equation}
as it can be verified in Eq. \ref{eq.leading}.
Indeed, upon substitution we get
\begin{equation} \label{eq.leading1}
    \log \bar{G}(h N^{y_H/d},T_c) = 
    N\frac{\delta}{c^{1/\delta}} \left(\frac{h}{\delta+1}\right)^{\frac{1+\delta}{\delta}} 
\end{equation}
yielding the expected leading magnetic singularity $\propto h^{d/y_H}$ of the free energy density of the model (Eq. \ref{eq.identity'}).

We propose to interpret the factor multiplying $N h^{d/y_H}$ on the r.h.s. of Eq. \ref{eq.leading1} as the amplitude of the singularity in $h$ of the free energy density $g(h,T_c)$ at $h=0$ (the coefficient $A$ in Eq. \ref{eq.identity'}).
While such an amplitude is generally expected to be non-universal \cite{Privman1984}, its estimation through the auxiliary function $\bar{G}$ suggests that it depends only on the universal exponent $\delta$ and the coefficient $c$ multiplying $\bar{m}^{\delta+1}$ in the exponential decay rate of $f$.
A direct verification that Eq. \ref{eq.leading1} provides precisely the amplitude of the leading singularity in $h$ of the free energy density in 2 or 3D is not possible.
However, our exact results for mean field interactions and for anomalous diffusion models provide strong indirect confirmation (see below).

The role played by the constant $c$ in the decay rate of $f$ acquires further importance if we look at the consequences of Eq. \ref{eq.leading1} for the rate function $I(m')$ in terms of which one expresses the large deviation principle for the magnetization density $m'=M/N$.
This principle can be formulated by saying that for large $N$ one has
\begin{equation} \label{eq.ldp}
    P_N(m'=M/N,T_c) \underset{N\to\infty}{\sim} e^{-NI(m')}.
\end{equation}
The validity of this large deviation principle is guaranteed here by G\"artner-Ellis theorem \cite{gartner1977on,ellis1984large}, since the condition $d/y_H > 1$ guarantees that $g(h,T_c)$ is differentiable for all $h$.
According to this theorem, we can obtain the rate function through a Legendre-Fenchel Transform of the SCGF as
\begin{equation} \label{eq.GE}
    I(m')=g(0,T_c) + \sup_{h>0} \left[ h m' -g(h,T_c)\right] .
\end{equation}
Such a supremum problem implies solving for $h$ the equation $m'=\partial_h g(h,T_c)$.
This can be easily done in the neighborhood of $h=0$, where the $h$ dependence of $g(h,T_c)-g(0,T_c)$ just reduces to the
leading singular term in Eq. \ref {eq.leading1}.
Substituting $h=c (1+\delta) (m')^\delta$ in the expression of which we need to compute the $\sup$ in Eq. \ref{eq.GE}, one obtains $I(m') = c (m')^{\delta+1}$, valid for $m' \to 0^+$.
This shows that the amplitude of the power law singularity of $I(m')$ is identical to the coefficient appearing in the exponential decay rate of $f(m)$ for large $|m|$.
This establishes a precise link between the singularity of the rate function $I(m')$ at its minimum $m'=0$, and the tail of the scaling function $f(m)$ for large $|m|$.
For critical Ising models, rate functions have been extensively studied, by numerical or renormalization group methods \cite{balog2022critical}.
However, so far, these studies did not establish for them general features related to universal critical exponents or to critical amplitudes. 

\medskip

\noindent \textit{Ising Mean-Field --}
An exact confirmation of the established above link between scaling and rate functions at criticality is possible in the case of an Ising model in which the spin-spin interaction is expressed in a mean field way as $\frac{J}{2 k_B T} M^2/N$.
This allows to write explicitly the magnetization probability $P_N(M,T)$ in zero external field as (see SM \cite{SM})
\begin{equation}
    P_N(M,T)\propto e^{N [\frac{1}{2}(M/N)^2\left(J/k_B T-1\right)-\frac{1}{12}(M/N)^4+\dots]} .
\end{equation}
The critical temperature is easily determined to be $T_c=J/k_B$ by cancellation of the quadratic term, implying a quartic rate function at criticality $I(m'=M/N)=-\frac{1}{12} (m')^4 + ....$.
This mean field rate function allows for an exact verification of the connection between its amplitude around $m'=0$ at criticality with the scaling function tail.

Since for the mean field case we should have $\delta=3$, we argue that a plausible scaled variable is $m=M/N^{3/4}$, which indeed cancels the asymptotic dependence on $N$ in the fourth order term.
So, for this model in which $d$ looses meaning, $N^{y_H/d}$ is replaced by $N^{3/4}$ in our basic Eq. \ref{eq.integral}.
The problem left is to determine $f(m)$.
Such a task was faced long ago by Ellis and Newman \cite{ellis1978fluctuationes,ellis1978limit} (see also \cite{cassandro1978critical}), with a derivation within a renormalization group strategy that proposed to identify the scaling function among a parameterized infinity of possible fixed point solutions which, however, does not include normalized PDFs.
In the Supplemental Material \cite{SM}, we present a consistent derivation of the scaling function for all $m\in[-\infty,+\infty]$ by showing that at criticality it exactly satisfies the stability condition
\begin{equation}
    f(m)= \frac{3^{1/4}\Gamma(1/4)}{2^{1/2}} f(2^{-1/4}m)^2
\end{equation}
under iterative doubling of the number of spins in the limit of infinite system, where $\Gamma(\cdot)$ is the complete Gamma function.
The unique solution of this fixed point equation is
\begin{equation} \label{eq.mfscaling}
    f(m)=\frac{\sqrt{2}}{3^{1/4}\Gamma(1/4)}e^{-\frac{1}{12}m^4}
\end{equation}
which is the (normalized) scaling function for the mean-field Ising model.
The exponential decay rate for $|m|\to\infty$ exactly coincides, both in exponent and in amplitude, with the leading term of $I(m')$ around $m'=0$, as predicted by our general argument.

\medskip

\noindent \textit{Central limit theorems --}
Another aim of the present investigation is to explore general, possibly universal, forms of limit theorems valid at criticality.
While the full knowledge of a scaling function would amount automatically to a limit theorem \cite{hilhorst2009central}, the existence of dependencies on boundary conditions of the critical scaling functions \cite{kaneda2001finite,antal2004} represents a manifest obstacle to the achievement of such generality. 
Our result concerning the asymptotic decay of scaling functions for large argument suggests validity of a weak form of generalized central limit theorem \cite{Touchette2009} for additive observables like the magnetization of the Ising model at criticality.
Indeed, the behavior we found for $I(m')$ around $m'=0$, suggests that one can approximate the PDF of $m'$ values not too deviating from $m'=0$ as
\begin{equation} \label{eq.CLT}
P(m') \sim e^{-N c (m')^{1+\delta}},
\end{equation}
depending only on $c$ and on the universal exponent $\delta$. 
As in the case of the standard central limit theorem, recovered here for $\delta=1$, one should expect that this approximation works well for magnetization values $M < \frac{1}{c} N^{\delta/(1+\delta)}$ if sub-leading contributions to $I(m')$ are of order sufficiently higher than $\delta+1$. 

Of key importance for the above derivations has been the possibility to extract the leading singular term of the free energy by performing the limit for $h N^{y_H/d} \to \infty$ (thus, at arbitrary nonzero $h$) of the integral in Eq. \ref{eq.integral}.
This term is obtained without having previously performed an $N \to \infty $ limit to extract the full $g(h,T_c)-g(0,T_c)$ in Eq. \ref{eq.identity'}.

\medskip

\noindent \textit{Anomalous Diffusion --}
Outside of equilibrium, large deviation theory can be used to describe the behavior of key extensive variables, such as the total particle transfer \cite{derrida2001free,derrida2002exact,derrida2007non,Touchette2009,derrida2011microscopic,wang2020large,pacheco2021large,chou2011non,mallick2022exact,baiesi2015role,teza2020rate} or entropy production \cite{lebowitz1999gallavotti,mallick2009some,teza2020exact}.
Dynamical large deviation functions in these contexts have been shown to exhibit singular behaviors, but typically related to first order transitions \cite{garrahan2007dynamical,lecomte2007thermodynamic,garrahan2009first-order,hedges2009dynamic,vaikuntanathan2014dynamic,baek2015singularities,nyawo2017minimal,whitelam2018large}.
Here we show that the derivation we outlined for the magnetization of the Ising model in equilibrium holds also for extensive variables exhibiting anomalous scaling out of equilibrium, ultimately leading to critical power-law singularities in the rate function.
Indeed, strong support to the validity of the above Ising scenario comes from the context of anomalous diffusion, in which the mean squared displacement of a diffusing particle grows as $\left< x^2 \right> \sim D t^{2\nu}$ with $\nu\neq 1/2$ and diffusion constant $D$ \cite{metzler2014anomalous,stella2023anomalous,stella2023universal}.
In these dynamical contexts, displacement plays the role of magnetization, while the system's extensivity is regulated by time in place of size.
Exact results can be obtained in the 1D Continuous Time Random Walk model (CTRW), in which a particle hops with some rate $r$ (in units $[t]^{-2\nu}$) to nearest neighboring sites on a lattice with spacing $L$.
The probability of observing a particle at a position $i$ of a 1D lattice evolves according to
\begin{equation}\label{eq:fractional_ME}
    [\partial_t^{2\nu} -2r] P_i(t)= r\left( P_{i-1}(t)+P_{i+1}(t) \right)
\end{equation}
where the fractional Caputo \cite{carpinteri2014fractals} derivative $\partial_t^{2\nu}$ accounts for the power-law tailed waiting time PDF $\omega(t)\sim t^{-1-2\nu}$ giving rise to subdiffusion \cite{barkai2000continuous}.
The generating function of displacement $G(\lambda,t)=\Sigma_i e^{\lambda i L} P_i(t)$ and the associated SCGF $\varepsilon(\lambda)=\lim_{t\to\infty} G(\lambda,t)/t$ can be evaluated exactly for this model (see \cite{SM,stella2023anomalous,teza2020exact}) providing the singular behavior
\begin{equation}\label{eq:SCGF_CTRW_leading_sing}
   \varepsilon(\lambda)=(rL^2)^{1/2\nu} |\lambda|^{1/\nu}+O(\lambda^{2+1/\nu})\ .
\end{equation}
Here the Laplace parameter $\lambda$ acts as the magnetic field $h$ in the Ising case, so that $\varepsilon(\lambda)$ can be regarded as an analog of $g(h,T_c)-g(0,T_c)$.

Taking the continuum limit, the lattice spacing goes to zero while $rL^2\to D$ stays constant and the probability distribution approaches the PDF $p(x,t)=\lim_{L\to 0}P_i(t)/L$ of the continuous displacement $x=iL$.
From Eq. \ref{eq:fractional_ME} we get that this PDF satisfies the fractional diffusion equation $\partial_t ^{2\nu} p(x,t)=D \partial^2_x p(x,t)$, whose solution can be expressed as $p(x,t)=M_\nu (x/t^\nu\sqrt{D})/(t^{\nu}\sqrt{D})$ with $M_\nu$ M-Wright function \cite{mainardi1994special,mainardi2010wright,schneider1989fractional,barkai2000continuous}.
This allows to conclude that the scaling function is $f(z/ \sqrt{D})= M_\nu(z/ \sqrt{D})$ with rescaled displacement $z=x/t^\nu$, so that we can write
\begin{equation}
    \bar{G}(\lambda,t) = \frac{1}{\sqrt{D}} \int dz\ e^{\lambda z t^\nu}
    M_\nu(z/\sqrt{D})\ .
\end{equation}
For large $|z|$ the M-Wright function is known exactly \cite{mainardi1994special} to have a stretched exponential decay
\begin{equation}
    M_\nu(z/\sqrt{D}) \sim \exp \left(-\frac{1-\nu}{\nu} \left( \nu |z|/ \sqrt{D} \right)^{\frac{1}{1-\nu}}\right). 
\end{equation}
This is consistent with $\delta =\nu/(1-\nu)$, equivalent to Eq. \ref{eq:delta_yH}, and expressing the Fisher relation for anomalous diffusion \cite{cecconi2022probability,stella2023anomalous,stella2023universal}.
Plugging $\delta$ and the $c$ coefficient implied by the last equation into the equivalent of Eq. \ref{eq.leading1}, and taking into account that $D=rL^2$, eventually yields the same leading singularity in $\lambda$ obtained in Eq. \ref{eq:SCGF_CTRW_leading_sing}. On the basis of this singularity one can
also argue the behavior of the rate function $I(v=x/t)$ around $|v|= 0$,
obtaining \cite{SM}
\begin{equation}
    I(v=x/t)=\frac{1-\nu}{\nu}\left( \frac{\nu |v|}{\sqrt{D}} \right)^{\frac{1}{1-\nu}}+o(v^{\frac{1}{1-\nu}}) .
\end{equation}

Summarizing, we have shown that plausible arguments allow to argue the exponential decay rate of the scaling function of a critical observable and to fully characterize the leading singularity of the  SCGF in terms of this rate. We did also establish the existence of a critical singularity of the rate function of large deviation theory at its minimum, with exponent and amplitude matching precisely those of this exponential decay rate. This singularity guarantees validity of a general form of central limit theorem for additive critical variables. The whole scenario must be expected to hold for both equilibrium and non-equilibrium variables obeying anomalous scaling with scaling function
possessing finite moments of all orders.
Altogether the results show the central role played by scaling function tails in setting up a large deviations approach to criticality. They also strongly support and extend, as far as meaning and implications are concerned, long standing unproved conjectures for equilibrium Ising criticality.

\begin{acknowledgments}
G. T. is supported by the Center for Statistical Mechanics at the Weizmann Institute of Science, the grant 662962 of the Simons foundation, the grants HALT and Hydrotronics of the EU Horizon 2020 program and the NSF-BSF grant 2020765. We thank Marzio Cassandro for advice and useful remarks in early stages of this project, as well as Giovanni Jona-Lasinio, Giovanni Gallavotti, Mehran Kardar, Hugo Touchette, David Mukamel, Oren Raz, and Amos Maritan for useful discussions and remarks.
\end{acknowledgments}


\bibliography{refs.bib}

\onecolumngrid

\bigskip
\hrule
\bigskip

\setcounter{equation}{0}
\renewcommand{\theequation}{S\arabic{equation}}

\begin{center}
    {\LARGE \textbf{Supplemental Material (SM)}}
\end{center}

In this supplemental material (SM) we discuss the details of the calculations and results presented in the main text.

\section{Magnetization Generating Function}
\noindent Let us take an Ising model system with reduced Hamiltonian at temperature $T$:
\begin{equation}
    \mathcal{H}(\vec{\sigma})=\frac{J}{k_B T}\sum_{\left< i,j \right>}\sigma_i \sigma_j +h \sum_{i=1}^{N} \sigma_i
\end{equation}
where $J>0$ is a ferromagnetic interaction modulating the nearest neighbors pairwise spin-spin interactions and $h$ is a dimensionless magnetic field. 
The partition function of the system at a temperature $T$ reads:
\begin{equation}
    Z_N(h,T)=\sum_{\vec{\sigma}} e^{\mathcal{H}(\vec{\sigma})}
\end{equation}
where the sum is performed over all $2^N$ spin configurations $\vec{\sigma}$ and the reduced Helmoltz Free energy consequently reads
\begin{equation}
    \mathcal{F}_N(h,T)=\log Z_N(h,T) = \log\left( \sum_{\vec{\sigma}} e^{\mathcal{H}(\vec{\sigma})} \right) .
\end{equation} 
Let us refer with $P_N(M,T)$ to the probability of observing a certain total magnetization $M=\sum_i \sigma_i$ in zero magnetic field $h=0$ at temperature $T$. Formally
\begin{equation}
    P_N(M,T)=\frac{e^{F_N(M,T)}}{Z_N(h=0,T)}
\end{equation}
where we have introduced the reduced Landau Free energy in zero magnetic field:
\begin{equation}
    F_N(M,T)= \log \left[ \sum_{\vec{\sigma} : \sum_i \sigma_i=M} e^{\frac{J}{k_B T}\sum_{\left< i,j \right>}\sigma_i \sigma_j} \right]
\end{equation}
and the sum is performed over all spin configurations $\vec{\sigma}$ yielding a total magnetization $M$.
Noting that the reduced Helmoltz free energy for any $h$ can be expressed in terms of the Landau free energy as
\begin{equation}
    \mathcal{F}_N(h,T)=\log \left[ \sum_M e^{F_N(M,T) +hM} \right]
\end{equation}
we get that the following identity holds
\begin{equation}
    G_N(h,T) :=\sum_M e^{h M} P_N(M,T)
    = \frac{\sum_M e^{F_N(M,T)+hM}}{Z_N(h=0,T)}
    = e^{\mathcal{F}_N(h,T)-\mathcal{F}_N(h=0,T)}
\end{equation}
where we introduced the generating function $G_N$ of the moments of the magnetization, which can be obtained upon derivation of $G_N$ with respect to the external magnetic field at $h=0$.

\section{Magnetization scaling function of an Ising Mean-Field ferromagnet}

\noindent The reduced Hamiltonian (at inverse temperature $\beta=1/k_B T$) for an Ising system in which every spin $\sigma_s=\pm 1$ is coupled to all other $N-1$ spins reads
\begin{eqnarray}
    \mathcal{H}(\vec{\sigma}_i)&=\frac{\beta J}{N}\sum_{s=1}^N\sigma_s \sum_{s'=s+1}^N \sigma_{s'} \\
    &=\frac{\beta J}{2N}\sum_{s=1}^N\sigma_s \sum_{s'\neq s} \sigma_{s'}
\end{eqnarray}
where $J>0$ represents the ferromagnetic interaction and $\vec{\sigma_i}$ one of the $2^N$ possible spin configurations of the system.
Note that in the following derivation we will use the inverse temperature $\beta$ in place of the temperature $T$.
The coupling strength is rescaled with the system size to maintain the energy extensive with the system size $N$.
Introducing the magnetization $M=\sum_{s=1}^N \sigma_s$ one can rewrite the Hamiltonian as an explicit function of $M$.
To do so, we note that $\sum_{s'\neq s} \sigma_{s'}= M-\sigma_{s}$ and substituting in the above expression we get
\begin{equation}
    \mathcal{H}(M)=\frac{\beta J}{2} \left( \frac{M^2}{N}-1 \right)
\end{equation}
In the derivation that follows, we will drop the constant term.

\subsection{Multiplicity}
\noindent A given magnetization $M$ corresponds to a fixed number $N_{\downarrow}$ of downwards pointing spins.
It's easy to see that $N_{\downarrow}(M)=N/2-M/2$.
For a given value of $M$ the multiplicity of the configurations amounts to
\begin{equation}
    \Omega_N(M)= {N \choose N_{\downarrow}(M)} = \frac{N!}{(\frac{N}{2}+\frac{M}{2})!(\frac{N}{2}-\frac{M}{2})!} .
\end{equation}
If we assume $N$ and $\frac{N\pm M}{2}$ large enough we can use Stirling's formula to approximate the logarithm of the multiplicity obtaining:
\begin{eqnarray}
	\log \Omega_N(M) &= -N \left[ \frac{1-M/N}{2} \log \left(\frac{1-M/N}{2}\right) + \frac{1+M/N}{2} \log \left( \frac{1+M/N}{2} \right)
	\right]+
	\log\sqrt{\frac{1}{\pi\frac{1-(M/N)^2}{2}}} +
	\log \frac{1}{\sqrt{N}} \\ \nonumber
	&=-N S(M/N) +
	\log\sqrt{\frac{1}{\pi\frac{1-(M/N)^2}{2}}} +
	\log \frac{1}{\sqrt{N}}
\end{eqnarray}
where we introduced the "entropy" function
\begin{equation}
S(x)= \frac{1-x}{2} \log \left(\frac{1-x}{2} \right) + \frac{1+x}{2} \log \left(\frac{1+x}{2} \right)
\end{equation}
defined for $x\in[-1;1]$.
For later use, we can already highlight its expansion around $x=0$, which reads:
\begin{equation}
    S(x)\sim -\log 2 + \frac{x^2}{2}+\frac{x^4}{12}+\frac{x^6}{30}+O(x^8)
\end{equation}
while the expansion of the square root term inside the logarithm reads $\sqrt{2/\pi}+x^2/\sqrt{2\pi}+O(x^4)$.

\subsection{Partition and probability functions}
\noindent Let us first define the partition function in zero magnetic field (from now on we set $J=1$):
\begin{equation}
    Z_N(\beta)=\sum_{M=-N}^N e^{\log \Omega_N(M)+\mathcal{H}(M)}
    =\sum_{M=-N}^N \binom{N}{\frac{N+M}{2}}e^{\frac{\beta M^2}{2N}}
\end{equation}
where the increments are of step 2.
This allows to define the exact probability of observing a given total magnetization $M$ as:

\begin{equation}
    P_N(M,\beta)=\frac{1}{Z_N}\binom{N}{\frac{N+M}{2}} e^{\frac{\beta M^2}{2N}}
\end{equation}
Now let us split the system into two sub-lattices each of size $N/2$, with total magnetizations $M_1$ and $M_2$.
Let us assume $M\geq0$ without any loss of generality.
The following equation relating the probabilities holds:
\begin{equation}
    P_N(M,\beta)=\frac{Z^2_{N/2}(\beta)}{Z_N(\beta)} \sum_{M_1=M-\frac{N}{2}}^\frac{N}{2}  P_{N/2}(M_1,\beta) P_{N/2}(M-M_1,\beta) e^{-\frac{\beta}{2N}(M-2 M_1)^2}
\end{equation}
In the case $M<0$ the sum spans the set $[-N/2,M+N/2]$.

\subsection{Ratio $Z_{N/2}^2/Z_N$}

\noindent The partition in terms of the entropy function $S(M/N)$ reads:

\begin{equation}
    Z_N(\beta)\sim \frac{1}{\sqrt{N}} \sum_{M=-N}^{+N}  e^{-N S(M/N) +\frac{\beta}{2N}M^2} 
    \sqrt{\frac{1}{\pi\frac{1-(M/N)^2}{2}}}
\end{equation}
Expanding the term in the sum around $M=0$ provides us with:
\begin{equation}
    Z_N(\beta)\sim \frac{2^{N}}{\sqrt{N}}\sqrt{\frac{2}{\pi}} \sum_{M=-N}^{+N} e^{\frac{M^2}{2N}(\beta-1)-\frac{M^4}{12N^3}}
\end{equation}
highlighting that at the critical temperature $\beta=\beta_c\equiv1$ the dominant term in the exponent is quartic in the magnetization.

\subsubsection{Critical case ($\beta=1$)}
\noindent At criticality, expressing the sum in terms of the rescaled magnetization $m=M/N^{3/4}$ (reminder, we also need to divide by a factor of two since the steps in the sum are of size 2) and taking the continuum limit yields
\begin{equation}
    Z_N(\beta_c)\sim \frac{2^{N}}{\sqrt{N}}\sqrt{\frac{2}{\pi}} \frac{N^{3/4}}{2} \int_{-N^{1/4}}^{+N^{1/4}} dm e^{-\frac{1}{12}m^4}
\end{equation}
where in the large size limit the integral domain can be extended to the whole real axis (the integrand function is peaked around the origin) providing the following constant value:
\begin{equation}
    \int_{-\infty}^{+\infty} dm e^{-\frac{1}{12}m^4}=\frac{3^{{1/4}}\Gamma(1/4)}{\sqrt{{2}}}
\end{equation}
Putting everything together we get:
\begin{equation}
    Z_N(\beta_c)\sim 2^N N^{1/4} \frac{3^{1/4}\Gamma(1/4)}{2\sqrt{\pi}}
\end{equation}
implying that the ratio of partition functions we wanted to estimate, in the limit of large $N$ reads:
\begin{equation}
    \frac{Z_{N/2}^2(\beta_c)}{Z_N(\beta_c)}\to_{N\to\infty} N^{1/4} \frac{3^{1/4}\Gamma(1/4)}{2\sqrt{2\pi}}
    \propto N^{1/4}
\end{equation}

\subsubsection{High temperature ($\beta<1$)}

\noindent In the case $\beta<1$, the leading term in the exponent is quadratic, implying a Gaussian scaling.
With the change of variable $z=M/N^{1/2}$ and taking the continuum limit, we can write the partition function in this case as:
\begin{eqnarray}
    Z_N(\beta)\sim \frac{2^{N}}{\sqrt{N}}\sqrt{\frac{2}{\pi}} \frac{N^{1/2}}{2} \int_{-N^{1/2}}^{+N^{1/2}} dz e^{\frac{\beta-1}{2}z^2} \to_{N \to \infty} 2^{N} \frac{1}{\sqrt{1-\beta}}
\end{eqnarray}
implying that the ratio of the partition functions in this case is the constant $1/\sqrt{1-\beta}$.

In the following section, we will address separately the critical and non-critical case to obtain a fixed point equation for the probability density function.

\begin{figure}
    \centering
    \includegraphics[width=0.75\linewidth]{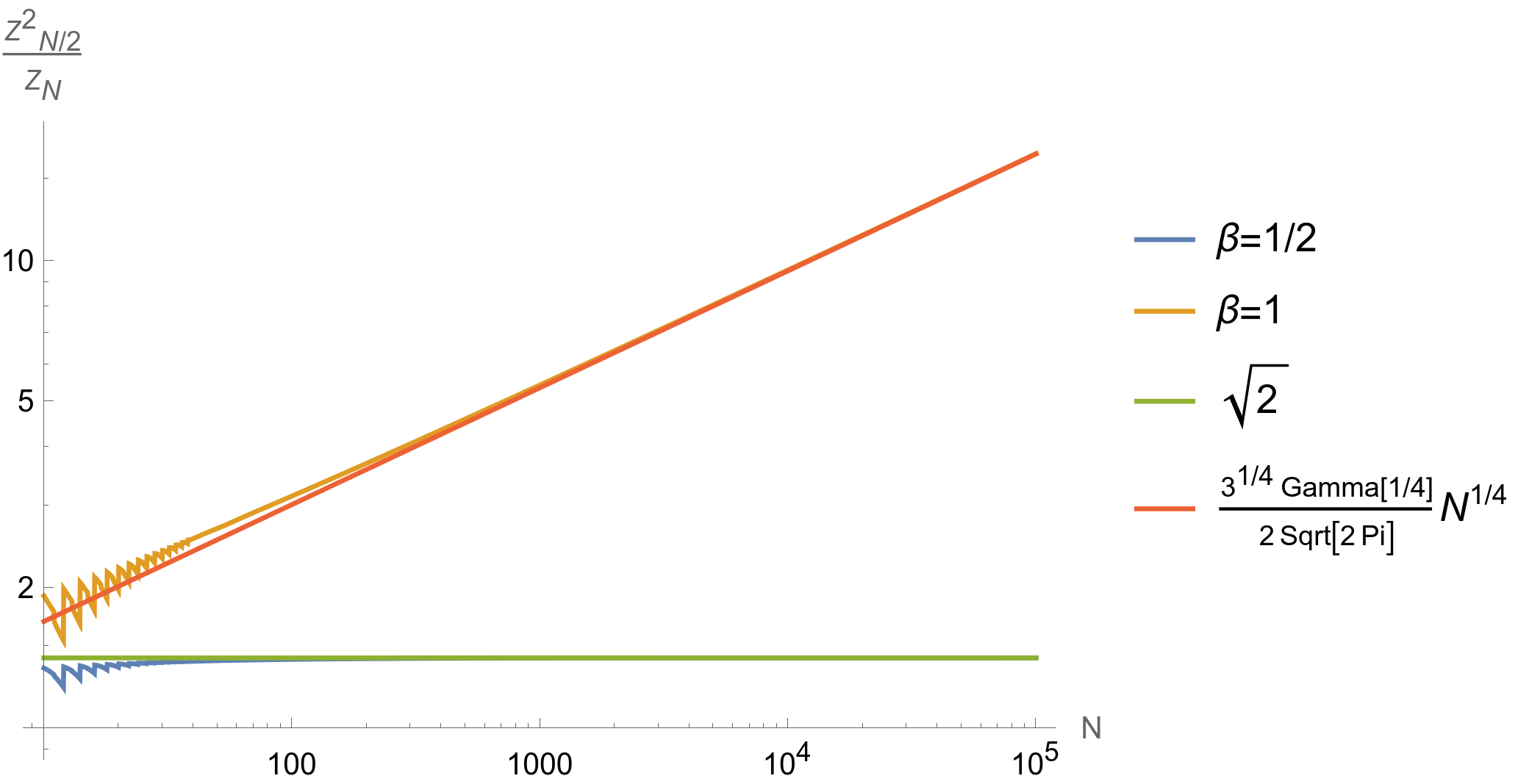}
    \caption{Ratio $Z_{N/2}^2/Z_N$ at criticality ($\beta=1$) and at high temperature $\beta>1$.
    The numerical calculation matches the predicted behavior, including the multiplying constants.}
    \label{fig:partitions}
\end{figure}

\subsection{Criticality ($\beta=1$)}

\noindent We assume that scaling holds in the limit of large $N$, namely that:
\begin{eqnarray}
    P_N(M,\beta_c)\to_{N\to\infty} 2 N^{-3/4} f(m)\ &\textrm{with:}\ m=M/N^{3/4} \\ \nonumber
    P_{N/2}(M_1,\beta_c)\to_{N\to\infty} 2 (N/2)^{-3/4} f(m_1)\ &\textrm{with:}\ m_1=M_1/(N/2)^{3/4} \\ \nonumber
    P_{N/2}(M-M_1,\beta_c)\to_{N\to\infty} 2 (N/2)^{-3/4} f(2^{3/4}m-m_1) \ &\textrm{with:}\  2^{3/4}m-m_1=\frac{M-M_1}{(N/2)^{3/4}} \\ \nonumber
    \sum_{M_1=M-\frac{N}{2}}^\frac{N}{2}\to_{N\to\infty}\left(\frac{N}{2}\right)^{3/4} \frac{1}{2} \int_{2^{3/4}m-(\frac{N}{2})^{1/4}}^{+(\frac{N}{2})^{1/4}} dm_1
\end{eqnarray}
We remind that the factors $2$ ($1/2$ in the integral) comes from the steps of length two for in the sum over $M$, consistently with the fact that we demand the scaling function to be normalized (i.e. $\int_{-\infty}^{+\infty}\ dm f(m)=1$).
This implies that for $m>0$ the relation
\begin{equation}
    \frac{2}{N^{3/4}}f(m)=\frac{Z^2_{N/2}(\beta_c)}{Z_N(\beta_c)} \left(\frac{N}{2}\right)^{3/4} \frac{2^2}{2} \int_{2^{3/4}m-(\frac{N}{2})^{1/4}}^{+(\frac{N}{2})^{1/4}} dm_1 \frac{f(m_1)f(2^{3/4}m-m_1)}{(N/2)^{3/2}}
    e^{-\frac{N^{1/2}}{2^{1/2}}(2^{-1/4}m-m_1)^2}
\end{equation}
holds asymptotically for large $N$, which, simplified, reads:
\begin{equation}
    f(m)=\frac{Z^2_{N/2}(\beta_c)}{Z_N(\beta_c)} 2^{3/4} \int_{2^{3/4}m-(\frac{N}{2})^{1/4}}^{+(\frac{N}{2})^{1/4}} dm_1 f(m_1)f(2^{3/4}m-m_1)
    e^{-\frac{N^{1/2}}{2^{1/2}}(2^{-1/4}m-m_1)^2}
\end{equation}

Substituting the ratio of the partition functions we evaluated previously at criticality we get:
\begin{equation}
   f(m)= N^{1/4} \frac{3^{1/4}\Gamma(1/4)}{2^{1/4}\sqrt{2\pi}} \int_{-\infty}^{+\infty} dm_1 f(m_1)f(2^{3/4}m-m_1)
    e^{-\frac{ N^{1/2}}{2^{1/2}}(2^{-1/4}m-m_1)^2}
\end{equation}
where we extend the integral bounds to $\pm\infty$ consistently with the large $N$ limit, which makes this equation holding true also for $m<0$.
With the change of variable $u=m_1-2^{-1/4}m$ we can rewrite the integral in a more symmetric form as:
\begin{equation}
   f(m)= \frac{3^{1/4}\Gamma(1/4)}{2^{1/2}} \int_{-\infty}^{+\infty} du f(2^{-1/4}m+u)f(2^{-1/4}m-u)
    \frac{N^{1/4}}{2^{1/4}\sqrt{\pi}}e^{-(N/2)^{1/2} u^2}
\end{equation}
where we have isolated the coefficients that will contribute to a Dirac-delta function in the integral.
This ultimately provides us with the fixed point equation:
\begin{equation}
    f(m)= \frac{3^{1/4}\Gamma(1/4)}{2^{1/2}} f(2^{-1/4}m)^2
\end{equation}
which can be shown to have as solution the probability density distribution
\begin{equation}
    f(m)=\frac{\sqrt{2}}{3^{1/4}\Gamma(1/4)}e^{-\frac{1}{12}m^4}.
\end{equation}
One can see that also the normalization in the fixed point equation is properly taken into account by integrating over $dm$ both sides of the equation
\begin{equation}
    f(m)dm= \frac{3^{1/4}\Gamma(1/4)}{2^{1/4}} f(2^{-1/4}m)^2 d(2^{-1/4} m)
\end{equation}
which yields the identity $1=1$.

\subsection{High temperature ($\beta<1$)}
\noindent We assume again that scaling holds in the limit of large $N$, only now with a different exponent. We have:
\begin{eqnarray}
    P_N(M,\beta)\to_{N\to\infty} 2 N^{-1/2} f(m)\ &\textrm{with:}\ m=M/N^{1/2} \\ \nonumber
    P_{N/2}(M_1,\beta)\to_{N\to\infty} 2 (N/2)^{-1/2} f(m_1)\ &\textrm{with:}\ m_1=M_1/(N/2)^{1/2} \\ \nonumber
    P_{N/2}(M-M_1,\beta)\to_{N\to\infty} 2 (N/2)^{-1/2} f(2^{1/2}m-m_1) \ &\textrm{with:}\  2^{1/2}m-m_1=\frac{M-M_1}{(N/2)^{1/2}} \\ \nonumber
    \sum_{M_1=M-\frac{N}{2}}^\frac{N}{2}\to_{N\to\infty}\left(\frac{N}{2}\right)^{1/2} \frac{1}{2} \int_{2^{1/2}m-(\frac{N}{2})^{1/2}}^{+(\frac{N}{2})^{1/2}} dm_1
\end{eqnarray}
This implies that for $m>0$ the relation
\begin{equation}
    \frac{2}{N^{1/2}}f(m,\beta)=\frac{Z^2_{N/2}(\beta)}{Z_N(\beta)} \left(\frac{N}{2}\right)^{1/2} \frac{2^2}{2} \int_{2^{1/2}m-(\frac{N}{2})^{1/2}}^{+(\frac{N}{2})^{1/2}} dm_1 \frac{f(m_1,\beta)f(2^{1/2}m-m_1,\beta)}{N/2}
    e^{-\beta(2^{-1/2}m-m_1)^2}
\end{equation}
holds asymptotically for large $N$, which, simplified, reads:
\begin{equation}
    f(m,\beta)=\frac{Z^2_{N/2}(\beta)}{Z_N(\beta)} 2^{1/2} \int_{2^{1/2}m-(\frac{N}{2})^{1/2}}^{+(\frac{N}{2})^{1/2}} dm_1 f(m_1,\beta)f(2^{1/2}m-m_1,\beta)
    e^{-\beta(2^{-1/2}m-m_1)^2}
\end{equation}
Substituting the ratio of the partition functions we evaluated previously for $\beta<1$ we get:
\begin{equation}
    f(m,\beta)=\frac{2^{1/2}}{\sqrt{1-\beta}} \int_{-\infty}^{+\infty} dm_1 f(m_1,\beta)f(2^{1/2}m-m_1,\beta)
    e^{-\beta(2^{-1/2}m-m_1)^2}
\end{equation}
where we extended the integral bounds to $\pm\infty$ consistently with the large $N$ limit, which makes this equation holding true also for $m<0$.
Introducing the characteristic function (Fourier transform) $\hat{f}(k,\beta)=\int dm e^{ikm}f(m,\beta)$ we can rewrite the above equation as:
\begin{equation}
    \hat{f}(k,\beta)= \frac{2^{1/2}}{\sqrt{1-\beta}} \int_{-\infty}^{+\infty} dm_1 f(m_1,\beta) e^{ik2^{-1/2}m_1}
    \int_{-\infty}^{+\infty} d \left( \frac{u}{2^{1/2}} \right) f(u,\beta) e^{ik2^{-1/2}u}e^{-\frac{\beta}{2}(u-m_1)^2}
\end{equation}
where we performed the change of variable $u=2^{1/2}m-m_1$.
It is easily seen that in the case $\beta=0$ we get
\begin{equation}
    \hat{f}(k,\beta=0)=\hat{f}(2^{-1/2}k,\beta=0)^2
\end{equation}
The solution is of the Gaussian family of functions $\hat{f}(k,\beta=0)=e^{-\alpha k^2}$ for any real $\alpha>0$.
Inverting to the real space, we get:
\begin{equation}
    f(m,\beta=0)=\frac{1}{2\sqrt{\alpha \pi}}e^{-\frac{m^2}{4\alpha}}
\end{equation}
This would correspond to the "weakly interacting variables" case outlined in \cite{cassandro1978critical}.


\section{Anomalous diffusion models}
\noindent A Continuous Time Random Walk (CTRW) model describes the evolution of a particle on a lattice with spacing $L$ through the fractional differential equation
\begin{equation}\label{eq:gen_ME}
    \partial^{2\nu}_t P_i(t)=r L^2 \frac{P_{i-1}(t)+P_{i+1}(t)-2 P_i(t)}{L^2}
\end{equation}
where $P_i(t)$ represents the probability of being on the $i$-th lattice site at a given time $t$, $r$ accounts for transition rate probability among nearest neighboring sites and the operator $\partial_t^{2\nu}$ (with $0<\nu<1$) represents the fractional Caputo derivative which has the following integral representation \cite{carpinteri2014fractals}:
\begin{equation}
      \partial_t^{2\nu} f(t) = \frac{1}{\Gamma(1-2\nu)}\int_0^t d\tau \frac{\partial_\tau f(\tau)}{(t-\tau)^{2\nu}}
\end{equation}
Introducing the generating function of displacement $G(\lambda,t)=\sum_i e^{\lambda iL} P_i(t)$ we get from Eq. \ref{eq:gen_ME} that it satisfies the equation
\begin{equation}
    \left[ \partial_t^{2\nu}-4r\sinh^2(\lambda L/2) \right] G(\lambda,t)=0
\end{equation}
which is exactly solved by the one parameter Mittag-Leffler function $E_{2\nu}(z)=\sum_{k=0}^{\infty} z^k/\Gamma(2\nu k1)$ (defined for every $z\in\mathbb{C}$ and $\nu>0$, see e.g. \cite{gorenflo2020mittag}) such that we can write
\begin{equation}
    G(\lambda,t)=E_{2\nu}\left( 4r\sinh^2(\lambda L/2) t^{2\nu}\right)
\end{equation}
The Mittag-Leffler function has an exponential (power-law) asymptotic behavior for positive (negative) arguments.
The argument of the function is positive for every $r>0$, so that the asymptotic representation for large times takes the exponential form
\begin{equation}
    G(\lambda,t)=\exp \left( (4r\sinh^2(\lambda L/2))^{1/2\nu} t +o(t) \right)
\end{equation}
which allows us to evaluate the exact SCGF as
\begin{equation}
    \varepsilon(\lambda)=\lim_{t\to\infty} \frac{1}{t} G(\lambda,t)=
    (4r\sinh^2(\lambda L/2))^{1/2\nu}=(rL^2)^{1/2\nu} |\lambda|^{1/\nu}+O(\lambda^{2+1/\nu})
\end{equation}
where we highlighted the leading singular character around $\lambda=0$ as presented in the main text.

The solution of Eq.\ref{eq:gen_ME} is expected to satisfy scaling for $t\to\infty$ and the equivalent of Eq. (1) in the main text can be written as
\begin{equation}
\frac{P_i(t)}{L} \rightarrow_{t\to\infty} \frac{t^{-\nu}}{(rL^2)^{1/2}}~ f \left( \frac{iL}{({rL^2})^{1/2}t^\nu} \right)
\end{equation}
The scaling function $f$ appearing in this equation can be obtained also
by performing a different limit in which $L\to 0$ and $r\to \infty$, keeping $rL^2=D$ fixed, in such away that $iL=x$ assumes the meaning of a continuous coordinate to be kept at finite values with $t$:
\begin{equation}
\lim_{L\to 0, r\to \infty, rL^2=D}~ \frac{P_i(t)}{L} = \frac{t^{-\nu}}{D^{1/2}}~ f\left(\frac{x}{D^{1/2}t^\nu} \right)   
\end{equation}
The possibility to determine $f$ is given by the circumstance that in this continuum limit Eq. \ref{eq:gen_ME} provides the fractional diffusion equation by simply recognizing on the r.h.s. a second order discrete central derivative of the limit probability density
\begin{equation}
    p(x,t)=\lim_{L\to 0, r\to\infty,rL^2=D}~\frac{1}{L}P_i(t) .
\end{equation}
Finally, recognizing on the r.h.s. of Eq. \ref{eq:gen_ME} a discrete second order central derivative, in this continuum limit we obtain the fractional diffusion equation
\begin{equation}\label{eq:FDDE}
    \partial_t^{2\nu} p(x,t) = D \partial_x^2 p(x,t) \ .
\end{equation}
By means of fractional calculus techniques \cite{carpinteri2014fractals}, one can show that this class of diffusion problems is exactly solved in terms of, e. g., M-Wright density function as $p(x,t)=M_\nu (x/t^\nu\sqrt{D})/(t^{\nu}\sqrt{D})$ where $M_{\nu}$ has the following representation \cite{mainardi1994special,mainardi2010wright,schneider1989fractional,barkai2000continuous}
\begin{equation}
    M_{\nu}(z)=\sum_{k=0}^{\infty}\frac{(-z)^k}{k! \Gamma(-\nu k +1-\nu)} .
\end{equation}
for every $z\in\mathbb{C}$.
Its asymptotic representation for large real valued $|z|$ can be expressed in terms of elementary functions as
\begin{equation}\label{eq:m_wright}
    M_{\nu}(z) \simeq a(\nu)|\nu z|^{\frac{\nu-1/2}{1-\nu}} e^{-b(\nu)|\nu z|^{1/(1-\nu)}}
\end{equation}
where the (positive) coefficients $a(\nu)$ and $b(\nu)$ have the form
\begin{equation}
    a(\nu)=\frac{1}{2\sqrt{2\pi(1-\nu)}},\ b(\nu)=\frac{1-\nu}{\nu}\ .
\end{equation}

\end{document}